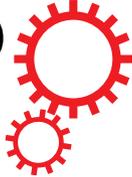

# Observation of strongly enhanced photoluminescence from inverted cone-shaped silicon nanostuctures

Sebastian W. Schmitt, George Sarau & Silke Christiansen

Silicon nanowires (SiNWs) attached to a wafer substrate are converted to inversely tapered silicon nanocones (SiNCs). After excitation with visible light, individual SiNCs show a 200-fold enhanced integral band-to-band luminescence as compared to a straight SiNW reference. Furthermore, the reverse taper is responsible for multifold emission peaks in addition to the relatively broad near-infrared (NIR) luminescence spectrum. A thorough numerical mode analysis reveals that unlike a SiNW the inverted SiNC sustains a multitude of leaky whispering gallery modes. The modes are unique to this geometry and they are characterized by a relatively high quality factor (Q ~ 1300) and a low mode volume ($0.2 < (\lambda/n_{eff})^3 < 4$). In addition they show a vertical out coupling of the optically excited NIR luminescence with a numerical aperture as low as 0.22. Estimated Purcell factors $F_p \propto Q/V_m$ of these modes can explain the enhanced luminescence in individual emission peaks as compared to the SiNW reference. Investigating the relation between the SiNC geometry and the mode formation leads to simple design rules that permit to control the number and wavelength of the hosted modes and therefore the luminescent emission peaks.

Since the band-to-band transition in silicon (Si) does not simultaneously obey the laws of energy and momentum conservation, spontaneous emission rates are low as compared to those of direct semiconductors and efficient Si based solid state lighting remains challenging[1]. At first, all-Si cm- to μm-sized waveguide lasers were realized by utilizing the optically more efficient stokes/anti-stokes transitions based on the stimulated Raman scattering effect[2], but these cannot be pumped electrically and only produce light of a particular quality[3]. Numerous other options to improve the efficiency of Si light emission have been demonstrated, most of which rely on quantum confinement in nanocrystals or quantum wells[4–6] and the introduction of luminescent defects or quantum dots by sophisticated materials engineering methods[7–9]. Furthermore, the confinement of spectrally matched optical modes can amplify spontaneous emission rates of electronic transitions in Si nanophotonic cavities. Hereby, the amplification is proportional to $Q/V_m$, where $Q$ is the quality factor of the cavity and $V_m$ is the volume of the resonant mode[10,11]. This so-called Purcell effect, could successfully be applied to improve the performance of Si-based light emitting devices (LEDs). Here, Si can act as an active or passive cavity material amplifying intrinsic band-to-band or defect transitions[8,12–16] or the transitions of added gain media like quantum dots or laser dyes[9,17].

While in previous studies an enhancement of Si (band-to-band or defect) luminescence could be shown in Si photonic crystals as compared to unstructured Si slabs, the presented study investigates an alternative approach. We show that the optically excited band-to-band luminescence of a straight SiNW attached to a Si wafer substrate is substantially increased by the introduction of a reverse taper. We find that the herby created inverse SiNC shows an about 200 times larger broadband NIR emission while in taper-induced spectral emission peaks the luminescent yield is multiplied by factors of up to ~670. A

Max Planck Institute for the Science of Light, Photonic Nanostructures, Günther-Scharowsky-Str. 1, 91058 Erlangen/Germany Helmholtz-Zentrum Berlin für Materialien und Energie, Institute Nanoarchitectures for Energy Conversion, Hahn-Meitner-Platz 1, 14109 Berlin/Germany. Correspondence and requests for materials should be addressed to S.W.S. (email: sebastian.schmitt@mpl.mpg.de or sebastian.schmitt@helmholtz-berlin.de)

thorough numerical mode analysis could reveal that the NIR light confinement in SiNCs is fundamentally different and superior as compared to that of SiNWs once both structures are supported by a Si substrate. Strong light confinement in the SiNCs takes place in multifold leaky whispering gallery modes (WGMs) with Q-factors of up to 1300 and low mode volumes $V_m$ of around $10^{-2}\,\mu m^3$. Due to the reverse conical geometry the SiNCs are characterized by vertical out-coupling of light from these WGMs so that numerical apertures (NA) as low as 0.22 could be shown. Estimated Purcell factors $F_p \propto Q/V_m$ of the modes in SiNCs can explain a more than 200-fold enhanced luminescence in individual emission peaks over the SiNW reference. Investigating the relation between the SiNC geometry and the mode formation leads to simple design rules that permit to control the number and wavelength of the hosted modes and therefore the luminescent emission peaks compared to the SiNW reference. Apart from its luminescent properties, the SiNC can be used as a highly controllable passive photonic nanocavity for Si-based nanostructure NIR light sources and lasers but may also be of interest for other fields of optoelectronics like optical sensing or as light concentrators for Si solar applications[18].

## Fabrication of the reversely tapered Si nanocones (SiNCs)

Figure 1a shows a scanning electron micrograph (SEM) of a representative SiNC (C4) and a reference SiNW (NW1). The structures are fabricated on a Si wafer (<100>, n-type/phosphorous, $1–5\,\Omega cm$) using polystyrene nanosphere lithography and cryogenic reactive ion etching (RIE) with an $SF_6$ and $O_2$ based plasma chemistry[18,19]. While the SiNW has a cylindrical shape with straight sidewalls, the RIE etched SiNC has the geometry of an inverted frustum. Both shapes can be tuned by modifying details of the RIE plasma etching receipts and the lithographic pattern. While the size of the polystyrene nanospheres determines the top diameter $D$, the etching time is proportional to the height $h$ of fabricated SiNWs and SiNCs. The sidewall taper (determined by the bottom diameter $d$ or the opening angle $\alpha$) is controlled by the $O_2$ concentration in the plasma which is responsible for the chemical sidewall passivation during the etching process. Post-processing of the SiNCs is carried out by thermal annealing (30 min, 500 °C) in $O_2$ atmosphere and a subsequent dip in hydrofluoric acid (HF, 5% in aqueous solution) to remove RIE induced fluorine contamination of the nanostructures' surfaces and remaining surface roughness. For electrical surface passivation and to form an optical cladding, 5 nm of $SiO_2$ are grown on the SiNC and SiNW surfaces by a second thermal annealing process in $O_2$ atmosphere (5 min, 500 °C). SEM images and exact dimensions of the different SiNCs and SiNWs fabricated for this study (SiNCs C1, C2, C3, C4, and a straight Si nanowire NW1 used for reference measurements) are presented in the supplementary information S1.

## Photoluminescence emission spectra

Figure 1b displays a scheme of the measurement setup to monitor the optically excited photoluminescence (PL) emission from the SiNCs. The excitation is carried out by a continuous wave (cw) laser with a wavelength of 660 nm. The normally incident laser light is tightly focused on the top of individual, well-separated, and substrate attached SiNCs and SiNWs by an objective (100×, NA 0.9). The emission spectra from these nanostructures are collected by the same objective in the backscattering configuration and are analyzed by a NIR spectrometer coupled to an InGaAs detector[20]. Figure 1c shows the numerically simulated cross sectional energy density in C4 and NW1 using excitation with 660 nm cw laser light. Even though the absorption maxima are located in different regions of the structures, the overall fraction of absorbed light $\eta_{abs}$ under the given excitation conditions (Gaussian beam profile with ~800 μm FWHM in –z direction, 660 nm cw) in both structures shows the same magnitude of about 40%. We start the discussion with the comparison of the emission spectra of C4 and NW1 excited with the 660 nm laser at a power of 1.28 mW displayed in Fig. 2a. The unique and truly different PL emission characteristics of the SiNC as compared to the SiNW becomes immediately evident, notwithstanding the fact that both structures have roughly the same volume and absorb the same amount of exciting light. While NW1 emits a broad, continuous PL spectrum distributed around the band edge of Si at (1100 nm/1.12 eV) the emission of C4 is substantially higher and exhibits multifold sharp emission peaks. The inset in Fig. 2a impressively displays the difference between the emission of C4 and NW1. For the SiNW, the InGaAs camera is hardly able to monitor the emitted NIR light, and only a few faint speckles are visible in the white circle (guide to the eye) centered in the position of NW1. In contrast, the emission of C4 appears as a confined bright spot in the center of the circle, indicating a highly collimated unidirectional beam emerging in z-direction (compare simulated emission characteristics of the modes in Fig. 4). The intensity ratio between the PL emission of C4 and NW1 is shown in Fig. 2b. While for individual wavelengths the emitted intensity of C4 exceeds the one of NW1 by factors as high as 673, the integral broadband emission was found to be about 204 times higher. Figure 2b shows the emission of all SiNCs fabricated for this study (compare C1–C4 in supplementary information S1) under an excitation power of 1.28 mW at 660 nm. It can be seen that a change of the SiNC geometry with respect to the angle of taper and the aspect ratio of gives rise to a different set of modes. As it will be shown in the mode analysis section for C4, the accommodated modes are specific to the geometry of the SiNCs. Measured Q-factors of the SiNC C1, C2, C3, and C4 ranged between 195–1308, 244–1047, 276–940, and 181–732, respectively (see the fitting procedure in supplementary information S2). In the following sections, the emission characteristics of the Si wafer surface attached SiNCs compared with the SiNWs will be explained by a numerical mode analysis and by the estimation of Purcell factors of the optical modes that could be found in the SiNCs.

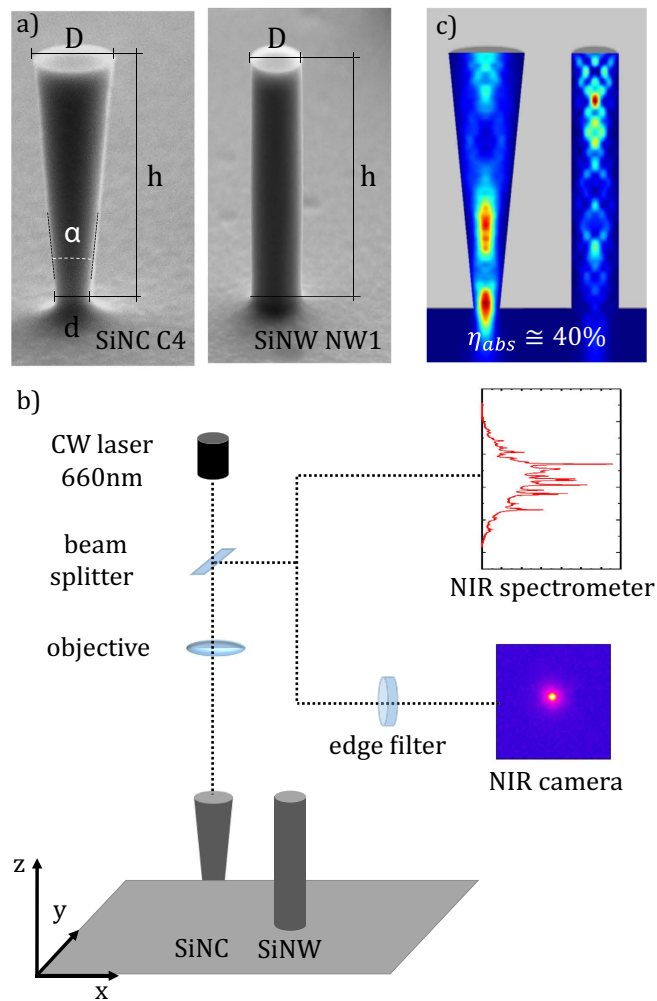

**Figure 1. SiNW and SiNC geometries and measurement setup.** (**a**) SEM of a typical SiNC (C4) and a reference SiNW (NW1) fabricated by RIE of a Si wafer with a shadow mask of silica nanospheres (oblique view with tilt angle of 70°). $D$, $d$, $h$, and $\alpha$ denote the top diameter, bottom diameter, height, and opening angle of the SiNC and SiNW, respectively (C4: $D = 941\,nm$, $d = 363\,nm$, $h = 2928\,nm$, $\alpha = 5.6°$, NW1: $D = 572\,nm$, $h = 2918\,nm$). (**b**) Schematic of the measurement setup: Through an objective ($100\times$, NA 0.9) the pump cw laser with wavelength of 660 nm is focused on the top surface of individual SiNCs or SiNWs. PL emission spectra are collected in the backscattering configuration using a VIS-NIR beam splitter (50/50) and are analyzed by a NIR spectrometer equipped with an InGaAs detector. The broadband emission of the SiNCs (SiNWs) is collected with an IR camera residing in the optical path after the beam splitter and an edge filter (750 nm cut on wavelength) to cut off the pump laser and Raman light. (**c**) Numerically simulated cross sectional energy density in C4 and NW1 under excitation with 660 nm cw laser light (red and blue color correspond to a high and low density respectively). The light fraction absorbed in both structures is $\eta_{abs}^{C4,NW1} \cong 40\%$.

## Mode Analysis

To understand the formation of strongly confined optical emission peaks in the SiNC that could not be found in the SiNW reference, a mode analysis using finite difference time domain simulations (FDTD solutions, Lumerical) was performed. For 660 nm pumping of a SiNC and a SiNW, the most laser light is absorbed at the maxima of the energy density as seen in Fig. 1c. Here, the highest spontaneous band-to-band emission from excited photo-carriers will emerge that, in turn, will give rise to NIR modes between 850 and 1250 nm that can be hosted by the SiNC and the SiNW. Therefore, for the mode

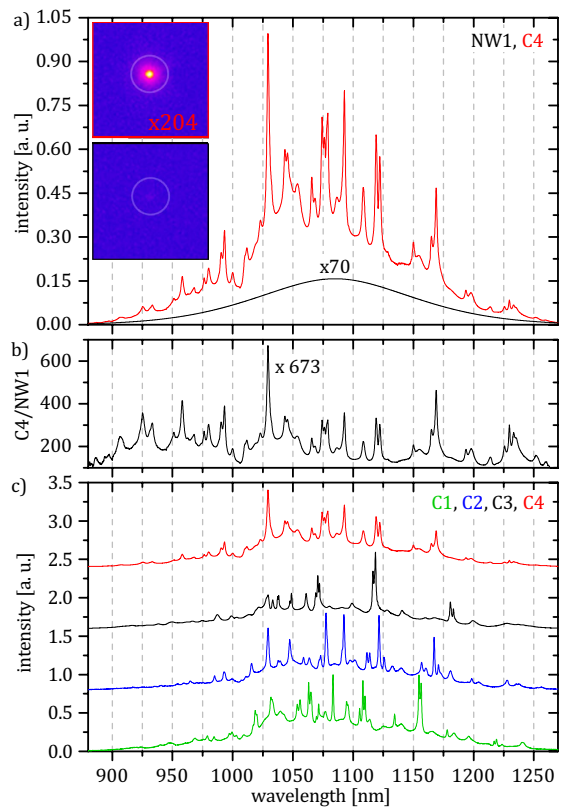

**Figure 2. PL emission spectra of SiNWs and SiNCs.** (**a**) Emission spectra from a SiNW (NW1) and a SiNC (C4) excited at 1.28 mW with a 660 nm cw laser. The signals are normalized to unity with respect to the highest peak. The SiNW spectrum is multiplied by a factor of 70 for better visualization. The insets show the broadband, far-field lasing emission of C4 and NW1 monitored by an InGaAs camera. The center of the white circles indicates the position of the SiNC and the SiNW, respectively. (**b**) Intensity ratio of the PL spectra (C4, NW1) shown in panel (**a**). (**c**) PL emission from SiNC with different geometries (C1, C2, C3, C4 in supplementary information S1) obtained using the same experimental conditions as described in panel (**a**). All signals are normalized to unity with respect to the highest peak intensity and shifted for reasons of clarity.

analysis, broadband dipole pulses (850–1250 nm) polarized in x and z direction ($E_x$, $E_z$) were excited in the maxima of the pump laser absorption in a SiNC with the geometry of C4 and a SiNW with the geometry of NW1. However, as is described in more detail in supplementary information S3, the SiNW attached to the wafer substrate is by far more optically 'leaky' than the SiNC, which is able to retain a higher amount of optical energy over a longer time span. This radiative energy is stored in optical modes. In the following, these modes, which are obviously unique to the SiNC geometry, will be analyzed in more detail. To find their spectral location, the radiative power emitted through the top facet of the SiNC was monitored in the numerical analysis. By plotting $P_z$ over the wavelength in Fig. 3a, a multitude of modes strongly radiating in the z direction can be identified, which is in good agreement with the experimental results in Fig. 2a. The peaks observed in the measured spectra show a good position coincidence with those in the simulated spectra. Moreover, numerically determined Q-factors of peaks in C4 ranging from 331 to 801 compare well with the measured ones ranging from 181 to 732 (for direct comparison see supplementary information S2 and S4). The amplitude of each resonance is not very meaningful, as it depends on the position of the sources used to excite the system and the position of the time monitors used to measure the response. However, in this case it is not necessary to calculate the absolute amplitude of the mode profile. The Q-factors and relative mode profile are the quantities of interest. Systematically smaller Q values in the measurements compared to the simulations can be attributed to surface roughness induced optical losses as e.g. already reported for Si micro disks or losses that are induced by free carrier absorption in the high injection regime (also compare next section)[21,22]. Furthermore, it can be seen that the set of excited modes is different for the two excitation polarizations ($E_x$, $E_z$). It turns out that all identified spectral radiation maxima correspond to WGMs that occur in discrete heights $h$ of the SiNC. Within the SiNCs, WGM positions are related to a circumference $U$ according to

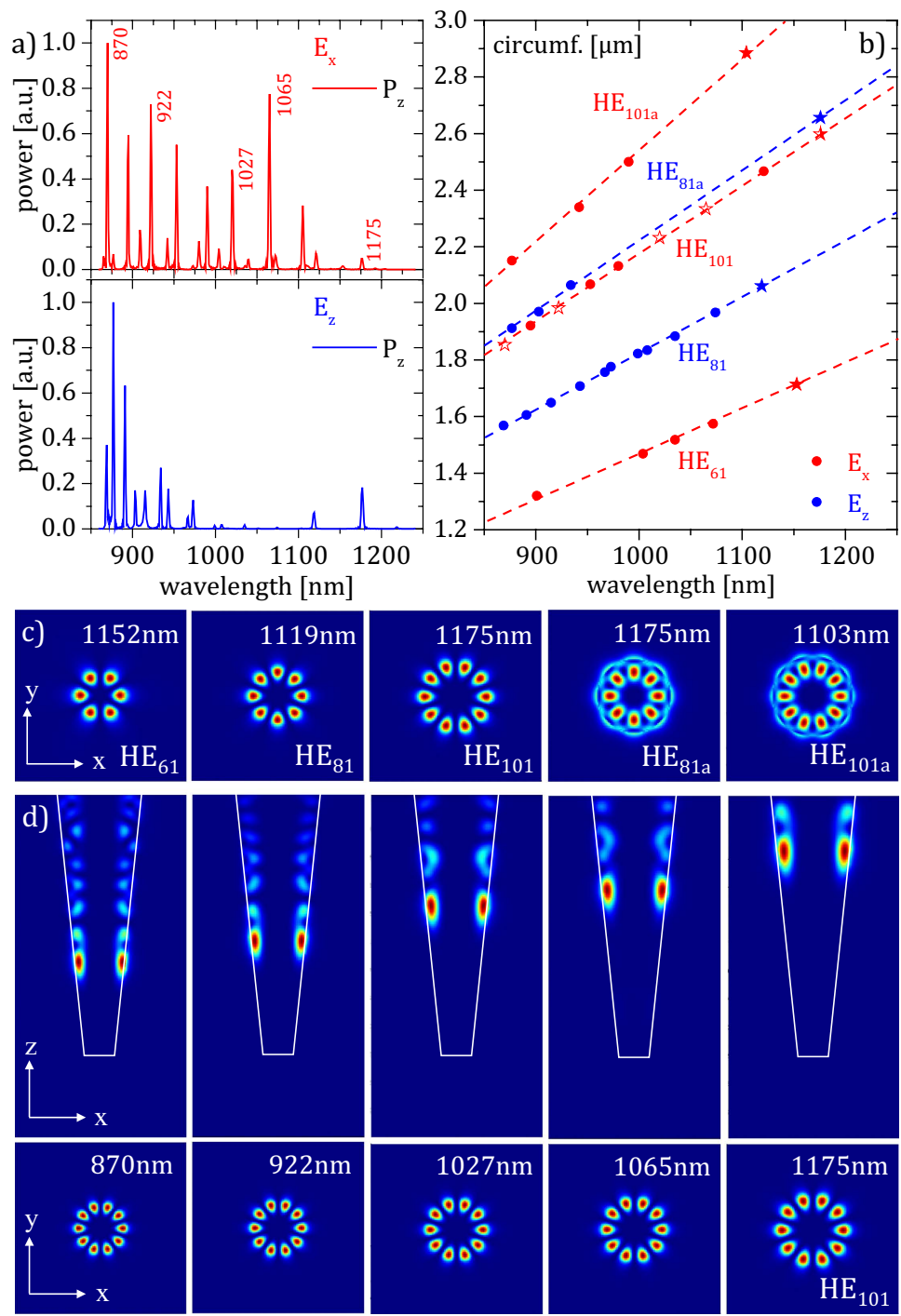

**Figure 3. Mode analysis.** (**a**) Radiative power from the modes through the top ($P_z$, arbitrary units) of the SiNC C4 plotted with respect to the wavelength. The upper (red) and lower (blue) graphs show modes excited by a dipole polarized along the x and z direction ($E_x$, $E_z$), respectively. The numbers indicate the peaks corresponding to the modes shown in (**d**). (**b**) Circumferences (heights) of the SiNC C4 hosting WGMs. Blue and red symbols correspond to WGMs excited by $E_x$ and $E_z$, respectively. Closed and open stars show WGMs visualized in (**c**) and (**d**). (**c**) Characteristic xy cross sections of the energy density for the $HE_{61}$, $HE_{81}$, $HE_{101}$, $HE_{81a}$, and $HE_{101a}$ WGMs. (**d**) xz and xy cross sections of the energy density for $HE_{101a}$ WGMs hosted at five discrete heights of the structure C4.

$U = \pi \cdot (2 \cdot h \cdot \tan\alpha - d)$ with $h$, $\alpha$, and $d$ are the characteristic dimensions of the SiNC specified in supplementary information S1. Plotting this circumference $U$ with respect to the spectral position of the maxima in Fig. 4b reveals the formation of five branches, along which WGMs align depending on the excitation by different polarizations ($E_x$, $E_z$). Each branch corresponds to a characteristic xy cross section of the energy density for the associated WGMs that in the following will be denoted as $HE_{61}$, $HE_{81}$, $HE_{101}$, $HE_{81a}$, and $HE_{101a}$, respectively. For the indexing of the modes see supplementary information S5. Figure 4c shows the typical xy cross sectional energy density for the WGMs in the five different branches with a wavelength of 1152 nm, 1119 nm, 1175 nm (2x), and 1103 nm. Note that $HE_{81a}$ and $HE_{101a}$ WGMs coincide at 1175 nm. xz and xy cross sections of the energy density of $HE_{101a}$ WGMs are shown in Fig. 3d. Surprisingly, the WGMs are hosted in discrete heights of the SiNC that in principle should offer a continuum of circumferences from bottom to top. The discretization can be explained by the fact that the high $Q$ WGMs found in the SiNC all show a long but much less intense 'leaking' branch towards the upper edge of the structure. As it can be seen in the xz cross sections of the energy density of the exemplary modes in Fig. 3d, this geometrically extended branch confines the modes to their discrete positions. The upward branch of the modes appears to be also responsible for the directional emission of the SiNC. Based on the numerical results for the 922 nm mode emission (see Fig. 3d), the NA of the SiNC C4 was estimated to be 0.22, which is about the NA of a typical optical fiber (see supplementary information S6), which implies the possibility of an efficient vertical NIR light coupling.

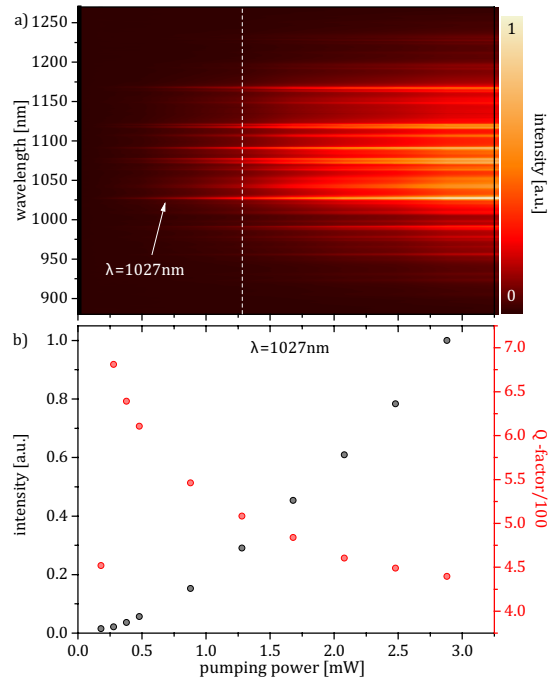

**Figure 4. Power dependence of PL emission.** (**a**) Spectral distribution of the normalized PL intensity of SiNC C4 as function of pump power (660 nm cw laser). The white dashed line indicates the position of the PL spectrum shown in Fig. 1(**a**). (**b**) Black dots show the normalized integrated PL intensity as a function of pump power for an individual peak in the spectrum (λ = 1027 nm) while red dots show the Q-factor which is determined by a fit of the associated peak (compare supplementary information S2).

## Power dependence and enhancement of photoluminescence

Figure 4a shows the spectral distribution of the normalized PL intensity of the SiNC C4 as function of pump power (660 nm cw laser). It can be seen that once the pump power exceeds 0.25 mW, sharp peaks begin to emerge from the PL background emission.

We focus our attention only on the most intense emission line of C4 at λ = 1027 nm. In Fig. 4b, it is shown that a steep rise of the line intensity with increasing pump power is correlated to a line broadening that leads to a decrease of the Q-factor. Note that for applied laser powers between 0.2 and 3 mW and an absorption of about 40% carrier generation rates $G$ in the SiNC C4 (or SiNW NW1, since both structures roughly have the same volume and absorption/Fig. 1a,c) range between $10^{20}$–$10^{21}$ cm$^{-3}$s$^{-1}$. This indicates that the Q-factor of the mode is limited by free-carrier absorption which is proportional to the concentration of free-carriers and therefore is supposed to be a dominant mechanism under the given excitation conditions. A further limitation of the Q-factor for higher pumping powers might be due to a temperature dependent line broadening, since pumping powers in the mW range will cause a substantial heating of the SiNC.

 6

At 1.28 mW pumping power, next to an overall enhancement factor of about 200 for the broadband emission, the most intense line in the C4 spectrum ($\lambda = 1027$ nm) shows an enhancement factor of 673 as compared to NW1 (Fig. 2a,b)! According to supplementary information S7, the spontaneous radiative emission power of light with wavelength $\lambda$ from a SiNC or SiNW can be written as

$$P_e = \eta_{out}\eta_{rad}\eta_{abs}\frac{hc}{\lambda}J \tag{1}$$

Here, $hc/\lambda$ is the energy of a photon with wavelength $\lambda$ and $\eta_{abs}$ is absorbed fraction of the photon flux $J$ injected by the pumping laser. Furthermore, $\eta_{out}$ represents the out coupling efficiency of the SiNC and SiNW for light of wavelength $\lambda$, while $\eta_{rad}$ is the radiative emission efficiency of the emitter including potential non-radiative carrier losses. We would like to show that an enhancement factor of 673 between the radiative emission power from the SiNC $P_e^{C4}$ and the SiNW $P_e^{NW1}$ cannot be explained by spontaneous emission as equated in (1) alone, and that there is a significant evidence for additional Purcell enhancement of the emission from the SiNC. For the given experimental conditions the absorbed power from the pump laser is the same for C4 and NW1, so looking at (1) the contribution of $\eta_{abs}(hc/\lambda)J$ to the emission enhancement can be neglected. Further it can be assumed that the radiative efficiency $\eta_{rad}$ of C4 and NW1 will essentially be comparable for the given experimental conditions, since it is merely dependent on material properties, material geometry and injection conditions (for details see supplementary information S8). This means the enhancement factor that can be explained by spontaneous emission, should mainly be related to a difference in the out coupling efficiency.

We numerically estimate the incoherent far field emission for C4 and NW1 and find $\eta_{out}^{C4} \cong 0.3$, $\eta_{out}^{NW1} \cong 0.1$. Accordingly, $P_e^{SiNC}/P_e^{SiNW} \cong 3$ leaving an unexplained enhancement factor of more than 200 for e.g. the $\lambda = 1027$ nm mode, which gives evidence of the Purcell effect acting on the spontaneous band-to-band emission in C4. Estimating the volume of hosted modes results in $V_m \cong 2-5\cdot10^{-2}\mu m^3$ (corresponding to $0.2 < \left(\lambda/n_{eff}\right)^3 < 4$/supplementary information S9). So using reasonable values for $Q \cong 500$ and a mode effective refractive index $n_{air} < n_{eff} < n_{Si}$, according to

$$F_p = \frac{3}{4\pi^2}\frac{Q}{V_m}\left(\frac{\lambda}{n_{eff}}\right)^3 \tag{2}$$

Purcell factors as high as 200 can be explained. Note that according to Fig. 4b for lower pumping energies and therefore higher Q-factors an even higher Purcell enhancement can be expected. Even though Q-factors of the presented leaky WGMs are relatively small as compared to the ones of e.g. Si photonic crystals or Si micro resonators, the high Purcell factor also results from the small mode volume in the structure ($F_p \propto Q/V_m$). Accordingly, as can be seen from supplementary information S9, of two modes with a comparable Q the one that circulates around a deeper orbit in the structure will show a potentially higher Purcell factor since it has a lower modal volume. To the best of our knowledge, this evaluation represents evidence for the highest enhancement of Si band-to-band luminescence in a photonic structure that has ever been reported[13]. Nevertheless, further measurements will be needed to find a direct prove of Purcell enhancement as e.g. the decrease of radiative lifetimes for the peak emission. Note that the study is not directly comparable to most previous work which discusses the enhancement of surface defect/dopant luminescence in Si photonic cavities, and not the band-to-band luminescence of the Si cavity itself[16,22–24].

## Design of customized inverted cone-shaped cavities

To numerically analyze the dependence of the photonic modes on the geometry of the SiNCs, broadband dipole pulses (850–1250 nm) polarized in x direction $(E_x)$ were excited in SiNCs of different geometrical shapes. Figure 5 displays the calculated radiative energy from the modes through the top facets $(P_z)$ of the SiNC C4 and three SiNCs with $D$ and $d$ similar to C4 and heights of $h$, $h/2$, $h/4$, and $h/8$, respectively, plotted with respect to the wavelength. It can be seen that the number of modes accommodated by the SiNC cavities continuously decreases with decreasing height. This finding is in good agreement with the results in Fig. 3b,d, which show the association of every mode with a discrete height in the structure. Consequently, once the height of the structure is reduced, only a lower number of confined modes is possible. This tendency can also be observed in Fig. 2c, where a lower height of the SiNCs corresponds to a lower number of emission peaks. The trend is not as clear as in the presented simulations, which span a much wider range of geometries. The lowest panel in Fig. 5 shows an additional variation of the SiNC bottom diameter from $d$ to $1.05\cdot d$ and $1.1\cdot d$, respectively. This change obviously is suitable to tune the spectral position of the modes in the structure. Here, a higher value for $d$ leads to modes with longer wavelengths, while the distance between the modes stays constant. Thus, the numerical investigation of the SiNCs shows that the photonic modes in the inverted cone-shaped nanocavities introduced in the present study can be customized by simple geometrical adjustments.

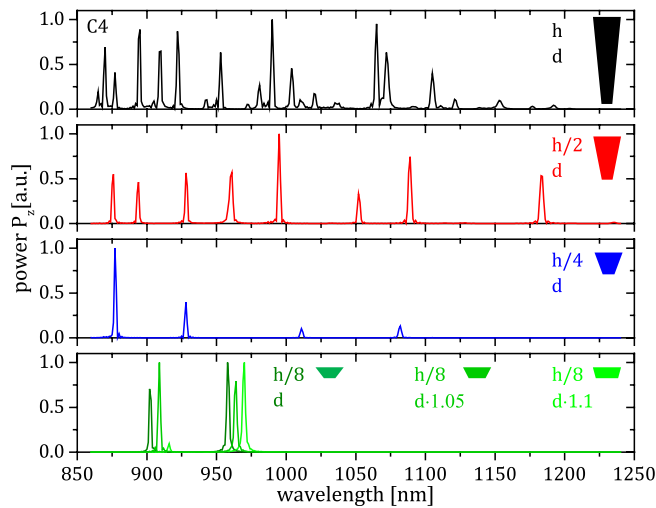

**Figure 5. Mode controllability.** Radiative energy from the high $Q$ modes through the top facets ($P_z$) of the SiNC C4 (top) and three SiNCs with $D$ and $d$ similar to C4 (see also Fig. 1) and reduced heights $h/2$, $h/4$, and $h/8$, respectively, plotted with respect to the wavelength. The lowest graph shows an additional variation of $d$ to $1.05 \cdot d$ and $1.1 \cdot d$. The graphs display modes excited by a dipole polarized along the x-direction ($E_x$) and are normalized to the highest peak in each spectrum.

## Conclusions

We introduce a new type of an all-Si photonic nanocavity in form of a reversely tapered Si nanocone (SiNC) attached to a Si substrate. The structure only slightly deviates from a silicon nanowire (SiNW) with straight sidewalls but in contrary to the SiNW exhibits a strong confinement of optically excited photoluminescence (PL) in multiple photonic whispering gallery modes (WGMs). Combined experimental and simulation results show that the formation of the unique and strongly confined WGMs is responsible for a strong enhancement and a directional out-coupling of Si band-to-band luminescence. Estimated Purcell factors of the WGMs are as high as 200 and can explain the PL enhancement of individual emission lines on top of the Si band-to-band spectrum. According to numerical simulations, the realization of a single-mode Si photonic resonator with an even higher light confinement appears possible.

The SiNC structure represents a very promising candidate for future applications as a CMOS compatible light source or as a building block for macroscopic Si-based light emitting diodes. Here, the introduction of a p-n junction into individual SiNCs can permit electrical driving[25], and estimated high Purcell factors could even encourage nano-lasing as soon as a coupling with an optical gain medium is realized[17]. Next to the use of the SiNC as a Si-based light source, its very sharp emission lines mark a starting point for novel optically detected sensing devices[26].

## Acknowledgements

The authors gratefully acknowledge financial support by the European Commission FP7: NMP priority-UNIVSEM (Nr. 280566), FIBLYS (Nr. 214042), and RODSOL (Nr. 227497); Health Priority - LCAOS (Nr. 258868) as well as by the German Research Foundation (DFG): FOR 1616. Further, the authors would like to acknowledge the funding of the Deutsche Forschungsgemeinschaft (DFG) through the Cluster of Excellence Engineering of Advanced Materials.

## Author Contributions

S.W.S. and G.S. conceived the idea of the inverted SiNCs. S.W.S. fabricated the SiNCs and explained the optical properties by numerical simulations. G.S. and S.W.S. performed the measurements and build the experimental setup. S.W.S. and G.S. analyzed the experimental data. S.W.S., G.S. and S.C. wrote the manuscript and discussed the work at all stages.

## Additional Information

**Supplementary information** accompanies this paper at http://www.nature.com/srep

**Competing financial interests:** The authors declare no competing financial interests.

**How to cite this article**: Schmitt, S. W. *et al.* Observation of strongly enhanced photoluminescence from inverted cone-shaped silicon nanostuctures. *Sci. Rep.* **5,** 17089; doi: 10.1038/srep17089 (2015).

# Observation of strongly enhanced photoluminescence from inverted cone-shaped silicon nanostructures

Sebastian W. Schmitt[*], George Sarau, Silke Christiansen

Max Planck Institute for the Science of Light, Photonic Nanostructures, Günther-Scharowsky-Str. 1, 91058 Erlangen / Germany

Helmholtz-Zentrum Berlin für Materialien und Energie, Institute of Nano-architectures for Energy Conversion, Hahn-Meitner-Platz 1, 14109 Berlin / Germany

[*]Corresponding author: sebastian.schmitt@mpl.mpg.de, sebastian.schmitt@helmholtz-berlin.de

***Supplementary information S1:*** SEM micrographs of four SiNCs (tilt 70°) and one SiNW (tilt 45°) fabricated by cryogenic RIE with $SF_6$ and $O_2$ chemistry (scale bar is 1µm). Masking is performed with silica spheres (diameter 1µm). Different shapes can be realized by the adjustment of $O_2$ concentration in the plasma that results in a different amount of under etching. The table gives the geometrical parameters of the produced nanostructures as determined by SEM and an image processing software.

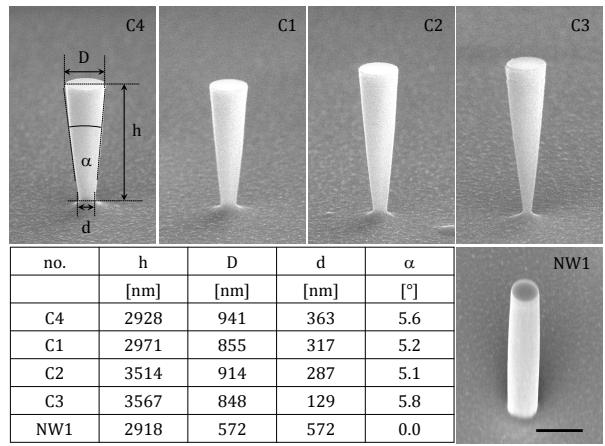

| no. | h | D | d | α |
|---|---|---|---|---|
| | [nm] | [nm] | [nm] | [°] |
| C4 | 2928 | 941 | 363 | 5.6 |
| C1 | 2971 | 855 | 317 | 5.2 |
| C2 | 3514 | 914 | 287 | 5.1 |
| C3 | 3567 | 848 | 129 | 5.8 |
| NW1 | 2918 | 572 | 572 | 0.0 |

***Supplementary information S2:*** The PL intensity $I$ from C1, C2, C3, and C4 was baseline corrected between 1000 and 1200nm (black line) and fitted (red line) by a sum of multiple Voigt peak profiles where each profile is given by

$$I(\lambda) = \int G(\tau)L(\lambda - \tau)\,d\tau \qquad (1)$$

with

$$G(\lambda) = \frac{e^{-\frac{\lambda^2}{2\sigma^2}}}{\sigma\sqrt{2\pi}} \qquad (2)$$

and

$$L(\lambda) = \frac{\gamma}{\pi(\lambda^2 - \gamma^2)} \qquad (3)$$

Here, $G$ and $L$ are a Gaussian and Lorentzian distribution with the widths $\sigma$ and $\gamma$, respectively. The spectral peak width $\Delta\lambda$ is then given by the approximation $\Delta\lambda = 1.1\gamma + \sqrt{0.9\gamma^2 + 8\ln 2\sigma^2}$. Accordingly the Q-factors were calculated as

$$Q = \frac{\lambda_o}{\Delta\lambda} \qquad (4)$$

where $\lambda_o$ is the spectral position of the peak maximum.

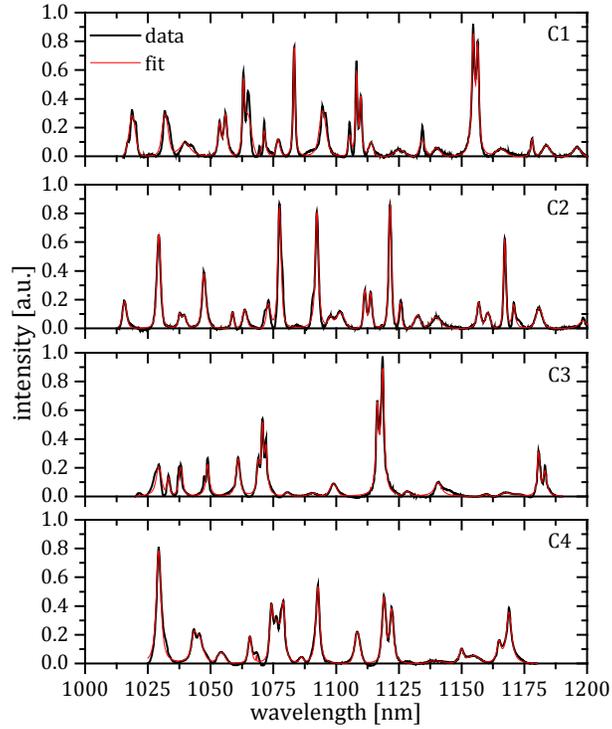

***Supplementary information S3:*** For the mode analysis, broadband dipole pulses (850-1250nm) polarized in x- and z- direction ($E_x$, $E_z$) were excited in the maxima of the pump laser absorption in a SiNC with the geometry of C4 and a SiNW with geometry NW1 (compare Fig. 1c). In a and b it can be seen that within the SiNW attached to the wafer, the pulse energy (proportional to $E^2$) decays much faster i.e. it is by far more optically 'leaky' than the SiNC, which is able to retain more optical energy over a longer time span. c and d show a Fourier transformation of the optical power emitted through the top facet of the SiNW and SiNC for t>600fs. While the spectra for the SiNC for different excitation still shows a strong emission in a multitude of sharp peaks, the emission of the SiNW is much weaker and only a few shallow peaks are visible in the spectrum. This is in good agreement to the PL spectra in Fig. 2a where in contrast to the SiNW the emission of the SiNC shows strong additional peaks.

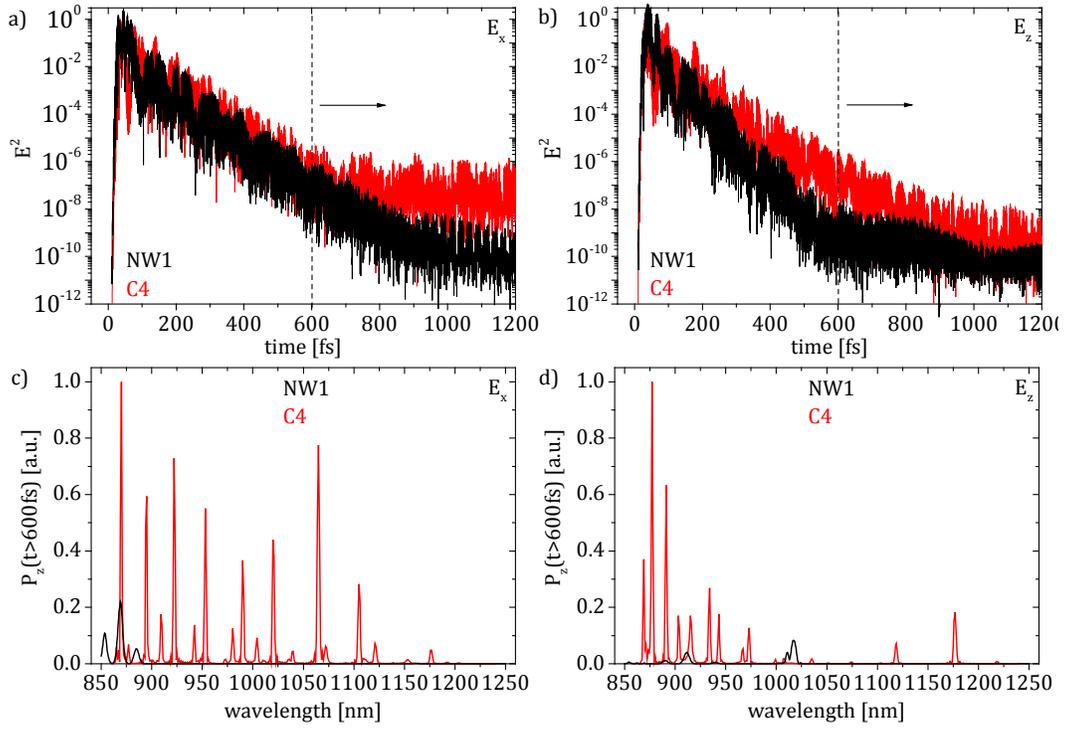

***Supplementary information S4:*** The figure shows the good agreement between the peak positions in the experimental emission spectrum of C4 and the positions of radiative energy maxima $P_z$ extracted from the numerical simulations. However, only about 63% of the peaks found experimentally are confirmed numerically. This can be explained by the fact that slight deviations in the geometry of the real SiNC C4 lead to the occurrence of additional modes and/or peak shifts that are not found by the simulations based on the ideal geometry of an inverted cone with dimensions as given in S1.

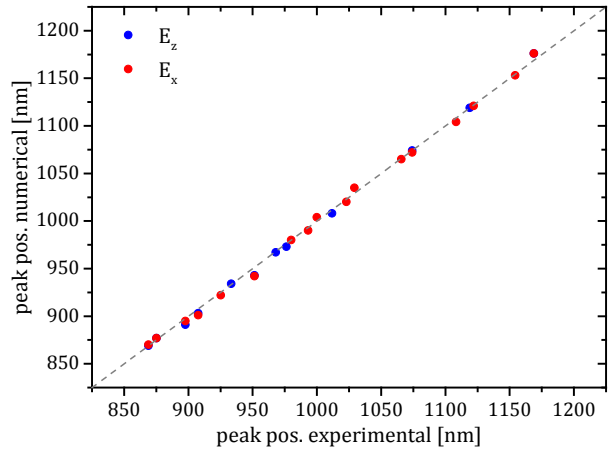

***Supplementary information S5:*** Figure 3c in the manuscript displays the xy cross sectional energy density for the WGMs at 1153nm, 1119nm, 1176nm (2x), and 1104nm. The exemplary chosen modes typically represent the branches $HE_{61}$, $HE_{81}$, $HE_{101}$, $HE_{81a}$, and $HE_{101a}$ identified in Fig. 3b. In $HE_{ij}$, the indices i and j correspond to the number of azimuthal and radial nodes, respectively[2]. With the additional index a, we distinguish the two appearance forms of $HE_{81}$ and $HE_{101}$.

***Supplementary information S6:*** The NA is estimated by a numerical analysis[3] of the far field radiation intensity of the modes in Fig. 3. We excite the SiNC C4 with a broadband dipole ($E_x$, as described in the main text) and monitor the radial distribution of the radiative energy in the z direction $P_z$ at a distance of $z_1$=20nm and $z_2$=200nm above the top facet (see the right scheme below). The graph shows the normalized radial distribution of the radiative intensity from the 922nm mode for the two distances. For simplified analysis, we use the double distance of the outer maxima, i.e. 80nm, as the broadening $b$ of the light cone between the distance of $z_1$=20nm and $z_2$=200nm away from the top facet. We apply

$$NA = n_{air} \cdot \sin(\tan^{-1}(b/\Delta z)) \qquad (5)$$

with $n_{air} \cong 1$ and $\Delta z = z_2 - z_1$ to estimate $NA = 0.22$.

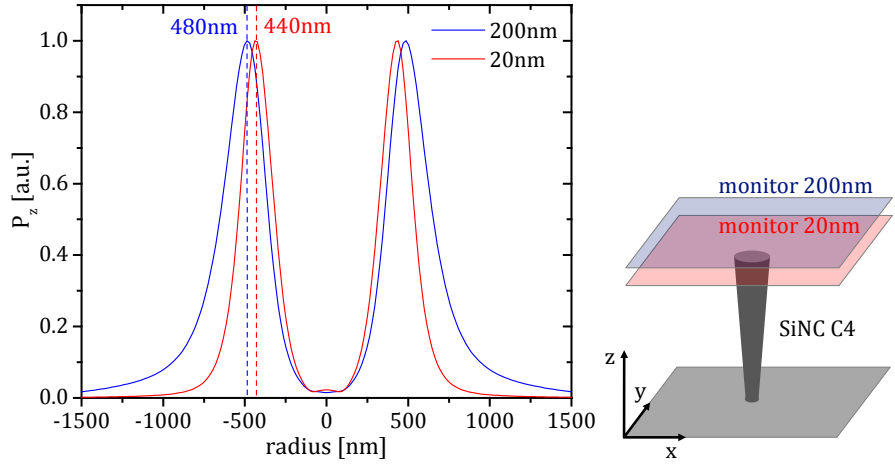

***Supplementary information S7:*** The temporal change of carriers $N$ in a unit volume of a solid state emitter is given by

$$\frac{dN}{dt} = G - R \qquad (6)$$

where $G$ is the generation rate and $R$ the sum of all carrier recombination processes. We rewrite $G$ as

$$G = \frac{\eta_{abs}J}{eV} \qquad (7)$$

with $\eta_{abs}$ representing the portion of the absorbed excitation intensity $J$, generating elementary charges $e$ in the volume $V$ of a SiNW or SiNC. $R$ can be expressed as the sum of all radiative and non-radiative recombination processes

$$R = R_{sp} + R_{nr} + R_{st} \qquad (8)$$

where $R_{sp}$ is the spontaneous (radiative) emission rate, $R_{nr}$ is the sum of no non-radiative recombination processes (and carrier leakage) and $R_{st}$ is the stimulated radiative emission. Assuming stationary conditions $dN/dt = 0$ and a low photon density leading to $R_{st} \cong 0$ the combination of (6), (7), and (8) leads to[4]

$$\frac{\eta_{abs}J}{eV} = R_{sp} + R_{nr} \qquad (9).$$

Using (9) and defining the radiative efficiency of the emitter as

$$\eta_{rad} = \frac{R_{sp}}{R_{sp} + R_{nr}} \qquad (10)$$

the internal spontaneous radiative emission power of light with wavelength $\lambda$, $P_i$

$$P_i = \frac{hc}{\lambda} V R_{sp} \qquad (11)$$

can be written as[4]

$$P_i = \eta_{rad}\eta_{abs}\frac{hc}{\lambda}J \qquad (12)$$

Here, $hc/\lambda$ is the energy of a photon with wavelength $\lambda$ and $\eta_{abs}$ is absorbed fraction of the photon flux $J$ injected by the pumping laser. To find the radiative emission $P_i$ emitted in direction of the analyzer, an out-coupling efficiency $\eta_{out}$ can be introduced, and accordingly[4]

$$P_e = \eta_{out}\eta_{rad}\eta_{abs}\frac{hc}{\lambda}J. \qquad (13)$$

**Supplementary information S8:** (10) can be rewritten as

$$\eta_{rad} = \frac{\tau_{sp}^{-1}}{\tau_{sp}^{-1} + \tau_{nr}^{-1}} \qquad (14)$$

in which $\tau_{nr}$ is the non-radiative recombination lifetime and $\tau_{sp}$ is the spontaneous emission lifetime. We calculate $\tau_{nr}$ for C4 an NW1 according to the approximation for SiNWs[5,6]

$$\frac{1}{\tau_{nr}} = \frac{1}{\tau_b} + \frac{4S}{d} \qquad (15)$$

where $S$ is the surface recombination velocity, $\tau_b$ is the bulk lifetime and $d$ the diameter of the SiNW. This approximation is valid for $S < \frac{2D}{d}$ where $D$ is the diffusion constant of carriers in Si. Since $S = 5 \cdot 10^2 \frac{cm}{s}$ for an oxidized Si surface[7] and $\frac{2D}{d} \cong 7 \cdot 10^3 \frac{cm}{s}$ with $D = 36 \frac{cm^2}{s}$ and $d \cong 500nm$, using (8) is justified in the presented case. The bulk lifetime of carriers in crystalline Si (n-type, 1-5$\Omega$cm) can range between $\tau_b \cong 10^{-4}s - 10^{-10}s$[8,9]. For a similar bulk material quality it is strongly decreasing at high injection levels (Auger-effect), so the value of $\tau_{nr}$ in (8) for a low injection is determined by the (in this case) very low surface recombination lifetime $\tau_{nr} = 4S/d \cong 10^{-8}s$, where in contrast for a very high injection it will be dominated the bulk Auger-recombination. Since C4 and NW have similar surface properties (SiO$_2$-passivation) and bulk material quality and their volume and surface/volume ratio is roughly the same, we can expect their $\tau_{nr}$ to be comparable for the same injection conditions. These are in fact given for the compared experimental spectra (Fig. 2a,b) that both have been acquired under excitation with an 1.28mW CW laser at 660nm, for which both structures absorb about 40% of the light incident at the top facet (Fig. 1c).

$\tau_{sp}$ can be calculated as[10]

$$\tau_{sp} = \frac{1}{N \cdot B} \qquad (16)$$

where the radiative recombination probability $B = 1.1 \cdot 10^{-14} \frac{cm^3}{s}$ and $N$ is the photo generated carrier density under optical pumping. Accordingly, $\tau_{sp}$ is dependent on intrinsic material properties and carrier injection, and therefore (as described above for $\tau_b$) will be comparable for C4 and NW1 under the given experimental conditions. This means that if a further Purcell enhanced emission can be neglected, $\eta_{rad}$ has about the same dimension for C4 and NW1.

***Supplementary information S9:*** The mode volume of all modes determined in C4 (Fig. 3b) was estimated numerically using

$$V_m = \int (E^2 > \frac{E^2}{2}) dV \quad (17)$$

We find $0.01\mu m^3 < V_m < 0.05\mu m^3$ intuitively a higher mode volume for a higher orbit of the leaky WGMs in the structure. For the mode at $\lambda$=1027nm we find $V_m \cong 0.01\mu m^3$ (see below a visualization of the torus containing the ½ of the mode optical energy).

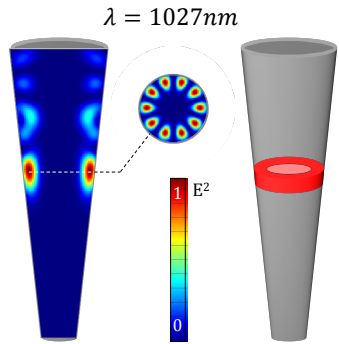